\documentclass[12pt]{iopart}

\usepackage{amsfonts}
\usepackage{mathrsfs}
\usepackage{amssymb}
\usepackage{color}
\usepackage[dvips]{graphicx}
\usepackage{wasysym}

\newcommand{\Z}{\mathbb{Z}}
\newcommand{\eq}{\begin{equation}}
\newcommand{\qe}{\end{equation}}
\newcommand{\eqa}{\begin{eqnarray}}
\newcommand{\qea}{\end{eqnarray}}

\newcommand{\halfWWW}{4.3cm}
\newcommand{\SLE}{\mathrm{SLE}}
\renewcommand{\Re}{\mathrm{Re}}
\renewcommand{\Im}{\mathrm{Im}}

\begin{document}
\title{Critical domain walls in the Ashkin-Teller model}
\author{M.~Caselle$^1$, S.~Lottini$^2$ and M.~A.~Rajabpour$^3$}
\address{$^1$ Dip.~di Fisica Teorica \& INFN, Universit\`a di Torino, Via P.~Giuria 1, 10125 Torino, Italy}
\ead{caselle@to.infn.it}
\address{$^2$ Johann Wolfgang Goethe Universit\"at, Max-von-Laue-Stra\ss{}e 1, 60438 Frankfurt am Main, Germany}
\ead{lottini@th.physik.uni-frankfurt.de}
\address{$^3$ SISSA and INFN, Sezione di Trieste, via Bonomea 265, 34136 Trieste, Italy}
\ead{rajabpour@sissa.it}

\begin{abstract}
We study the fractal properties of interfaces in the 2d Ashkin-Teller model. 
The fractal dimension of the symmetric interfaces is calculated along 
the critical line of the model in the interval between the Ising and 
the four-states Potts models. Using Schramm's formula for crossing probabilities 
we show that such interfaces can not be related to the simple $\SLE_{\kappa}$, except 
for the Ising point. The same calculation on non-symmetric interfaces is performed
at the four-states Potts model: the fractal dimension is compatible with the result 
coming from Schramm's formula, and we expect a simple $\SLE_{\kappa}$ in this case.
\end{abstract}

\vspace{2pc}

\maketitle

\section{\label{sec:introduction}Introduction}
Studying critical systems has been in the interest of physicists for at
least five decades. In two dimensions many exact methods to study critical
systems were invented by physicists and mathematicians, such as the Yang-Baxter equation \cite{baxter}, conformal field
theories (CFT) \cite{Di_Francesco}, and recently Schramm-Loewner
evolution ($\SLE$) \cite{schramm}. Most of the critical systems in two
dimensions can be formulated in terms of fluctuating loops: for example, in
the most familiar case, i.~e.~the Ising model, they simply correspond to
domain walls between regions of
opposite magnetization. The $\SLE$ is based on direct investigation of
these critical loops and is based on probability techniques
(for a review, see \cite{Kager}). According to $\SLE$ all of the conformally
invariant curves in two dimensions can be parametrized by the $\SLE$
drift $\kappa$, related to the fractal dimension of the curves via
$D=1+\frac{\kappa}{8}$.

One of the most interesting statistical systems to study from the
$\SLE$ point of view is the Ashkin-Teller (AT) model \cite{Ashkin}. This model
is interesting at least from two points of view: first, it has a rich
phase diagram including a critical line, and, secondly, it has
an interesting physical realization as Selenium adsorbed on the Ni(100)
surface \cite{bak}. Following the above motivations we investigate
the critical loops in this model on the critical line.

The definition of the Ashkin-Teller model on an arbitrary graph is
as follows: on each vertex $i$ of the graph lives a field $s_i$ with
values $s_i = e^{-i\frac{\pi}{2} q_i}$, where $q_i=0,1,2,3$.
Then the partition function of the model is 
\mbox{$Z = \sum_{\{s_{i}\}}\prod_{<ij>}W(s_i,s_j)$}, where
\eq
    W(s_i,s_j) = \big(1+x_{1}s_{i}^{*}s_{j}+x_{2}s_{i}^{*2}s_{j}^{2}+x_{1}s_{i}^{*3}s_{j}^{3}\big) \;,
	\label{AT_Z}
\qe
with the product over pairs of adjacent sites. The above
partition function is apparently $\Z_4$ symmetric and for
$x_{1}=x_{2}$ reduces to the four-states Potts model. The model was
solved exactly on the square lattice by mapping it to the $6$-vertex
model and some of the critical exponents were found by mapping the
model to Solid-on-Solid (SOS) model (\cite{wu_and_lin, 
Kadanoff_and_Brown, nienhuis2}; see also \cite{Kohmoto_den_nijs} for the phase
diagram of the anisotropic case). The self--dual Ashkin-Teller model
on the square lattice is described by the line $x_{2}+2x_{1}=1$;
moreover, it is exactly solvable all along this line. Some special
points along this line are in the universality class of well known
models: the point $x_{1}=x_{2}=\frac{1}{3}$  is in the universality
class of the four-states Potts model. The model is critical for all
points on the self--dual line with $x_{1}\geq \frac{1}{3}$. At the
point $x_{1}=\frac{sin(\pi/16)}{sin(3\pi/16)}$, called
Fateev-Zamolodchikov (FZ) point, the model is fully integrable and
can be described by $\Z_4$ parafermionic CFT \cite{ZF}. By an easy
change of variables, see (\ref{eq:z_ab}), one can show that the
above partition function corresponds to the Hamiltonian of two
coupled Ising models which decouple for $x_2=x_1^2$, so for
$x_{1}=\sqrt{2}-1$ we have the Ising universality class. Since the critical properties of
the fluctuating curves at the above special points on the critical
line are known, one can use them to extrapolate the results to
other points.

From the CFT point of view it is well-known that  the field theory
describing the Ashkin-Teller model at the critical
line is the $c=1$ orbifold conformal field theory \cite{yang,Saleur,
vongehlen_and_rittenberg}.
This CFT comes from compactifying the free bosonic field theory
to a circle with radius $r$ and then requiring a $\Z_2$ symmetry
for the bosonic fields (for a review of this conformal field
theory see \cite{Ginsparg, Schellekens}). To fix the notation take
the free field theory with action $S=\frac{1}{4\pi}\int\partial\phi
\bar{\partial}\phi$ and then compactify the bosonic field on a
circle of radius $r$. The `electromagnetic' conformal spectrum of the model is
\begin{equation}
  X_{em} = \frac{e^2}{2r^2} + \frac{r^2 m^2}{2} \,,
\end{equation}
where $e$ and $m$ are the electric and magnetic charges respectively. 
The action is invariant under $\phi\rightarrow-\phi$. To get an orbifold 
CFT we should project all the
operator content of the circle theory to the operators which
respect this symmetry, however, this theory cannot be modular
invariant without introducing some twisted operators which
come from the discrete $\Z_2$ symmetry of the model. In the case of $\Z_2$
orbifold CFT we have four operators, two of them have conformal
weights $\frac{1}{16}$, and we will call them $\sigma$, which is
reminiscent of the spin operator in Ising model. The other two have
conformal weights $\frac{9}{16}$ and are called $\tau$. This theory
is conformally invariant for all real values of $r$ and by changing
the radius we can change continuously the critical exponents of
the  model except the twist operators which are invariant under a change of
orbifold radius. On the square lattice  we have the following
equality between the radius and the coupling of Ashkin-Teller model on the
critical line \cite{yang,Ginsparg}:
\eq
    \label{AT_R}
    \sin\Big(\frac{\pi r^{2}}{8}\Big)=\frac{1}{2}\Big(\frac{1}{x_{1}}-1\Big) \;\;
\qe
(an analogous relation for 
practical calculations on the triangular lattice is given in the last section). 
A lattice-invariant way to write the above equation, which is also useful in
numerical calculations, is
\eq
    \label{AT_thermal_exponent}
    \frac{1}{\nu}=2-\frac{2}{r^{2}}\;\;,
\qe where $\nu$ is the thermal exponent. Some well-known points
are the following: $r=2$ is the four-states Potts model, $r=\sqrt{3}$ describes 
$\Z_4$ parafermionic CFT (see \cite{ZF}), $r=\sqrt{2}$ is in the Ising 
universality class -- indeed it is a pair of decoupled Ising models -- and 
finally $r=1$ is in the universality class of the $XY$ model.

There are different possibilities to define critical curves in the
Ashkin-Teller model \cite{gamsacardy07,ps1,ps2,ps3,IR}.    In the spin 
representation one can think about the domain walls between one 
spin and the other three, or the domain walls between
two definite spins and the other two. It is obvious that at the
Ising point the latter ones, if properly chosen, give the domain walls of the Ising
model, so we will mainly focus  on a ``symmetric'' choice recovering
standard Ising interfaces on one of the two underlying $\Z_2$
systems, except at the Potts point where we consider also the former choice. We should stress here that there 
is no direct connection between these interfaces and the domain walls of the orbifold Gaussian free field theory. 
The contour lines of Gaussian free field theory are related to a different kind of interfaces discussed in \cite{IR}. 
As was recently proven in \cite{chelkak}, the $\SLE$ drift of the ``symmetric'' choice at the Ising point
is $\kappa=3$. At the FZ point there is a prediction by Santachiara \cite{Santachiara} 
stating that the critical curves are related to $\kappa=\frac{10}{3}$. Finally, at the four-states
Potts model the common belief is that the $\SLE$ drift is $\kappa=4$. Except the Ising case 
(for which now there is a mathematical proof) there is no definite argument showing the specific connections 
of the above predictions to particular interfaces. In other words we do not know exactly to which specific 
interfaces the above predictions, coming from scaling-limit arguments, should be associated.
In this paper we  
examine the above predictions indirectly and we also systematically study the ``symmetric'' choice 
of the domain walls. Our method is based on calculating the fractal dimension of the introduced 
interfaces and comparing the results with those coming from checking Schramm's 
formula for the crossing probability. The most direct way to see the conformal invariance of the interfaces is 
using the inverse Loewner equation, finding the corresponding drift and comparing it with the Brownian motion for a normal 
$\SLE_{\kappa}$ and with more complicated stochastic processes for the $\SLE_{\kappa,\rho,\rho}$. However, it is difficult 
to get results with high accuracy by working directly with the $\SLE$
equation: a numerically more accessible way is to rule out the possibility of having $\SLE_{\kappa}$ by just calculating the 
crossing probability and checking it against Schramm's formula.

The structure of the paper is as follows:
in the next section we fix the notations and give the outline of the numerical 
procedure; in the third section we calculate the fractal dimension of the 
interfaces in the region between the Ising and the four-states Potts models; the fourth section 
is a check of Schramm's formula for the introduced interfaces; finally, in the last section we 
summarize our findings and also make some comments about possible exact 
formulas for describing them. 

\section{\label{sec:definitions}Definitions and procedure}
The numerical part of this work is conveniently expressed in the coupling space $\beta,\alpha$, where
\eqa
    Z = \sum_{\{\sigma_r;\tau_r\}}\prod_{<ij>}e^{S_{ij}} \;\;;\\
    S_{ij} = \beta(\sigma_i\sigma_{j}+\tau_i\tau_{j})+\alpha(\sigma_i\sigma_{j}\tau_i\tau_{j})\;\;.
    \label{eq:z_ab}
\qea
Here, $\sigma$ and $\tau$ are two Ising variables; the correspondence is $s_i =
\frac{e^{-i\frac{\pi}{4}}}{\sqrt{2}}(\sigma_i+i\tau_i)$, so that
\eq
    x_1 = \frac{e^{2\beta}-e^{-2\beta}}{e^{2\beta}+e^{-2\beta}+2e^{-2\alpha}} \;,\;
    x_2 = \frac{e^{2\beta}+e^{-2\beta} - 2e^{-2\alpha}}{e^{2\beta}+e^{-2\beta}+2e^{-2\alpha}}\;.
    \label{eq:ab_to_x1x2}
\qe
In this language the Ising model has $\alpha=0$, and the Potts
line is $\alpha=\beta$; the numerical work was carried on in the
triangular lattice because in this case an interface can be defined
in a unique and natural way (the dual lattice possesses only
three-links joints); on this geometry, the critical line is 
described by \cite{temperley1979}
\eq
	e^{4\alpha}(e^{4\beta}-1)=2\;\;.
\qe

Two kinds of interfaces are considered in the following, called $12|34$ and
$1|234$. Here, numbers denote the four possible spin states 
$q_i$, 
and the interface is realized by imposing that the spins on the either half of the 
border can assume only some of them. In the two-Ising notation above, one can identify, modulo
$Z_4$ transformations, $1=(\sigma=+,\tau=+)$, $2=(+,-)$, $3=(-,-)$ and $4=(-,+)$.

Most of this work deals with the $12|34$ interface, which amounts 
to imposing the boundary condition $\sigma=+1$
on one half of the system's border and $\sigma=-1$ on the other,
while leaving $\tau$ free.
This allows for a fast cluster-based update strategy,
which is an adaptation of the Swendsen-Wang prescription applied
alternatively to the Fortuin-Kasteleyn clusters of the $\sigma$ and $\tau$
sublattices, thus helping preventing the critical slowing 
down \cite{z4_gauge_dualAT} (in practice, the fixed boundaries are represented by
an additional layer of $\sigma$ spins which participate in forming the clusters but are not
allowed to be flipped).
For the $1|234$ interface, instead, we used a local Metropolis accept/reject
algorithm: the interface was induced by restricting the pool of possible
spins on the system boundary to $\{1\}$ and $\{2,3,4\}$ on the two halves.
In this case the simulations were
performed on Graphics Processing Units, so to exploit their
huge parallelisation capability, with the CUDA programming libraries.

The strategy to identify the fractal dimension $D$ associated to a particular
point $(\beta,\alpha)$ along the critical line of the AT model went as follows:
for a variety of system sizes $L\times L$ the associated interface length $S(L)$
was measured over a large number of configurations; this function is expected
to have the leading-order behaviour
\eq
	S(L) = f_{1p}(L) = a L^D\;\;.
	\label{eq:fracdim_order1}
\qe

Given a spin configuration, the associated interface length $S$ is obtained 
by recolouring isolated sign-clusters to leave only two regions, and then counting
the links connecting spins of opposite sign (much in the same way as described in
\cite{gamsacardy07}, with the difference that instead of an actual recolouring
we build the simply connected tree of neighboring sign-clusters in order to track,
among all opposite-colour interfaces, the desired one).

\begin{figure}
    \begin{center}
    \includegraphics[width=5.5cm]{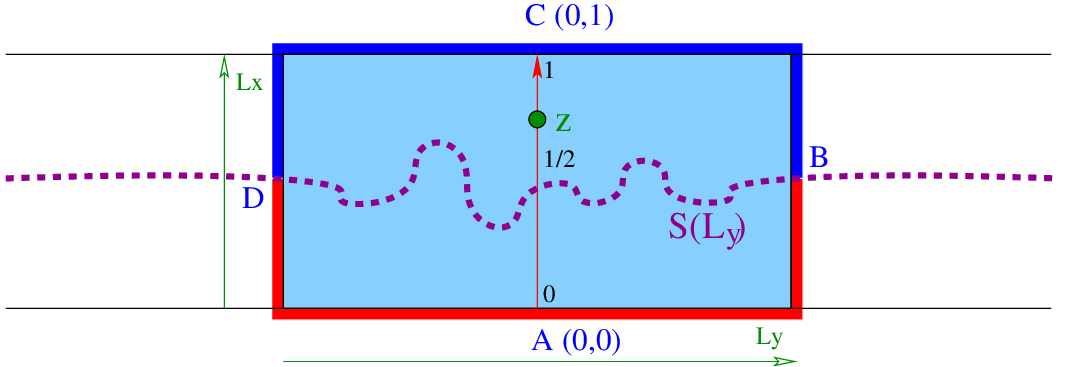}
	\hspace{0.8cm}
    \includegraphics[width=\halfWWW]{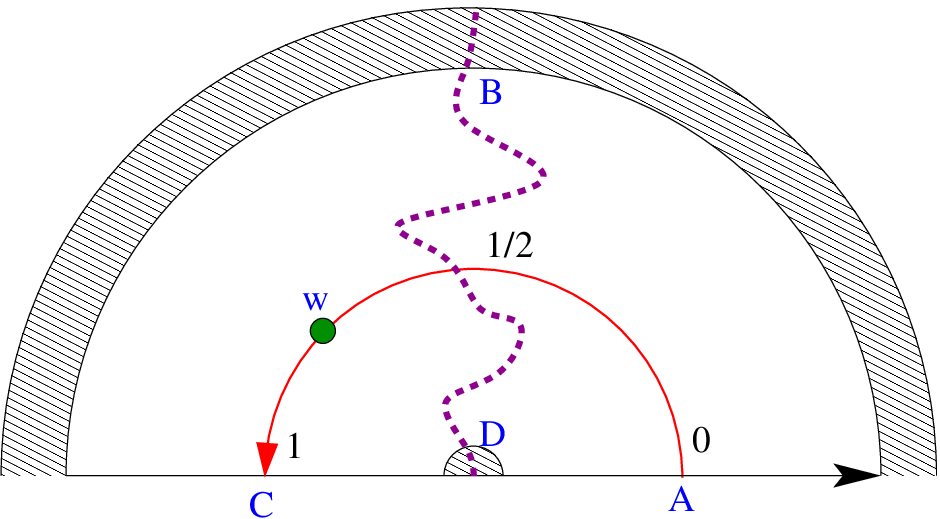}
    \caption{Geometric setting for the right-passage probability. The purple dashed
	line represents the interface; measurements are taken along the red line. The conformal mapping
	between the rectangle and the upper half-plane is such that the shaded areas of the latter,
	namely $|w|\gtrless\exp(\pm\pi \ell)$, are left out.}
    \label{fig:mapping-sketch}
    \end{center}
\end{figure}

Once the interface has been obtained, and all sites have been identified as lying on the
left or on the right of it, collecting data for the crossing probability 
is straightforward. What we measure is $F(x)$, where $0\leq x\leq 1$ is a coordinate
along a segment perpendicular to the interface and located halfway between the two opposite sides
of the system, and $F$ is the probability that the point at $x$ lies above the interface.
The theoretical Schramm prediction for this function -- provided the interface is described
by a $\SLE_\kappa$ -- is formulated on the upper half-plane, with the interface connecting
points $0$ and $+i\infty$; the original rectangular system is an approximation of the infinite
strip with width 1 in the complex plane, where the point on the segment has complex coordinates $z(0,x)$,
which are subsequently mapped in a conformal way to the half-plane via $w=e^{\pi z}$ (figure~\ref{fig:mapping-sketch}).
In order to have a good approximation of the whole half-plane, then, it is necessary to
work with elongated systems, that is, we examined aspect ratios $\ell = L_y/L_x$ from 1 to 5.
Plugging those transformation into the Schramm formula \cite{schramm_paper_original_formula},
\eq
	P_\kappa(w) = \frac{1}{2} - \frac{\Gamma(\frac{4}{\kappa})t}{\sqrt{\pi}\Gamma(\frac{8-\kappa}{2\kappa})}
		\;_2F_1\Big(\frac{1}{2},\frac{4}{\kappa},\frac{3}{2};-t^2\Big)\;\;,
	\label{eq:starting_schramm}
\qe
where $_2F_1$ is the hypergeometric function and $t = \frac{\Re(w)}{\Im(w)}$, 
we get the following formula for practical applications
\eq
	P_\kappa(x) = \frac{1}{2} - \frac{\Gamma(\frac{4}{\kappa})\cos(\pi x)}{\sqrt{\pi}\Gamma(\frac{8-\kappa}{2\kappa})}\;
		_2F_1\Big(\frac{1}{2},\frac{3\kappa-8}{2\kappa},\frac{3}{2};\cos^2(\pi x)\Big)\;\;,
	\label{eq:final-schramm}	
\qe

\section{\label{sec:df}Fractal dimension from the interface length}
The subleading corrections to (\ref{eq:fracdim_order1}) are still not
completely clear, and their direct numerical investigation is very hard,
so we tried several possible functional forms:
if the resulting values of $D$ appear to agree to some extent, one can
use this pool of results to assess the systematic errors involved in
determining the fractal dimension. Beyond the leading-order formula
quoted above, we considered a power-law correction,
\eq
	f_{2p}(L) = a_1 L^D + a_2 L^w \;\;;\;\;w < D\;\;,
\qe
(with $w$ either left free or fixed to $1$), and a logarithmic subleading term
\eq
	f_{ln}(L) = a_1 L^D + c \log(L/L_0)\;\;.
\qe
In practice, we fit the measured $S(L)$ to these functions in a range
$L_\mathrm{min} \leq L \leq L_\mathrm{max}$ by varying $L_\mathrm{min}$
looking for plateaux of stable parameters and acceptable $\chi^2/ndf$.
We also tried, at the Potts point, other functional forms inspired by 
the RG arguments in \cite{Aharony_Asikainen_analytical_corrections}, 
but they had to be discarded in favour of the above.

For the $12|34$ interface, we examined six points on the triangular-lattice
critical line: the three exactly known
Potts, F--Z and Ising universality classes (4P, FZ and I respectively),
plus three others points, labelled B, C and D. We also collected data for
the $1|234$ interface at the Potts point. In table \ref{table:points_studied}, 
we characterize all points by reporting 
the correlation exponent $\nu$, which is known as a function
of the couplings \cite{mohammad_steo_isoradial}.

\begin{table}
\caption{\label{table:points_studied} Couplings and correlation exponent for the six 
points that we considered on the critical line.}
\begin{indented}
\item[]\begin{tabular}{@{}llll}
\br
	Point name & $\beta$ & $\alpha$ & $\nu$ \\
\mr
	4P & $\frac{\log(2)}{4}$ & $\frac{\log(2)}{4}$ & $\frac{2}{3}$ \\
	D & $0.174007$ & $0.1718484$ & $0.678870$ \\
	FZ & $\frac{1}{4}\log\big( 1+\frac{2}{\sqrt{3}} \big)$ & $\frac{\log{3}}{8}$ & $\frac{3}{4}$ \\
	B & $0.216942$ & $0.0924784$ & $0.823489$ \\
	C & $0.242110$ & $0.0505546$ & $0.897473$ \\
	I & $\frac{\log{3}}{4}$ & $0$ & $1$ \\
\br
\end{tabular}
\end{indented}
\end{table}

For each point (and each choice of interface), we measured $S(L)$ 
for various tens of values of system size up to $L\sim 1000$--$2000$.
The statistics employed was approximately of half a million 
configurations for each $L$ and each point in phase space.

\begin{figure}
    \begin{center}
    \includegraphics[width=5.26cm,angle=270]{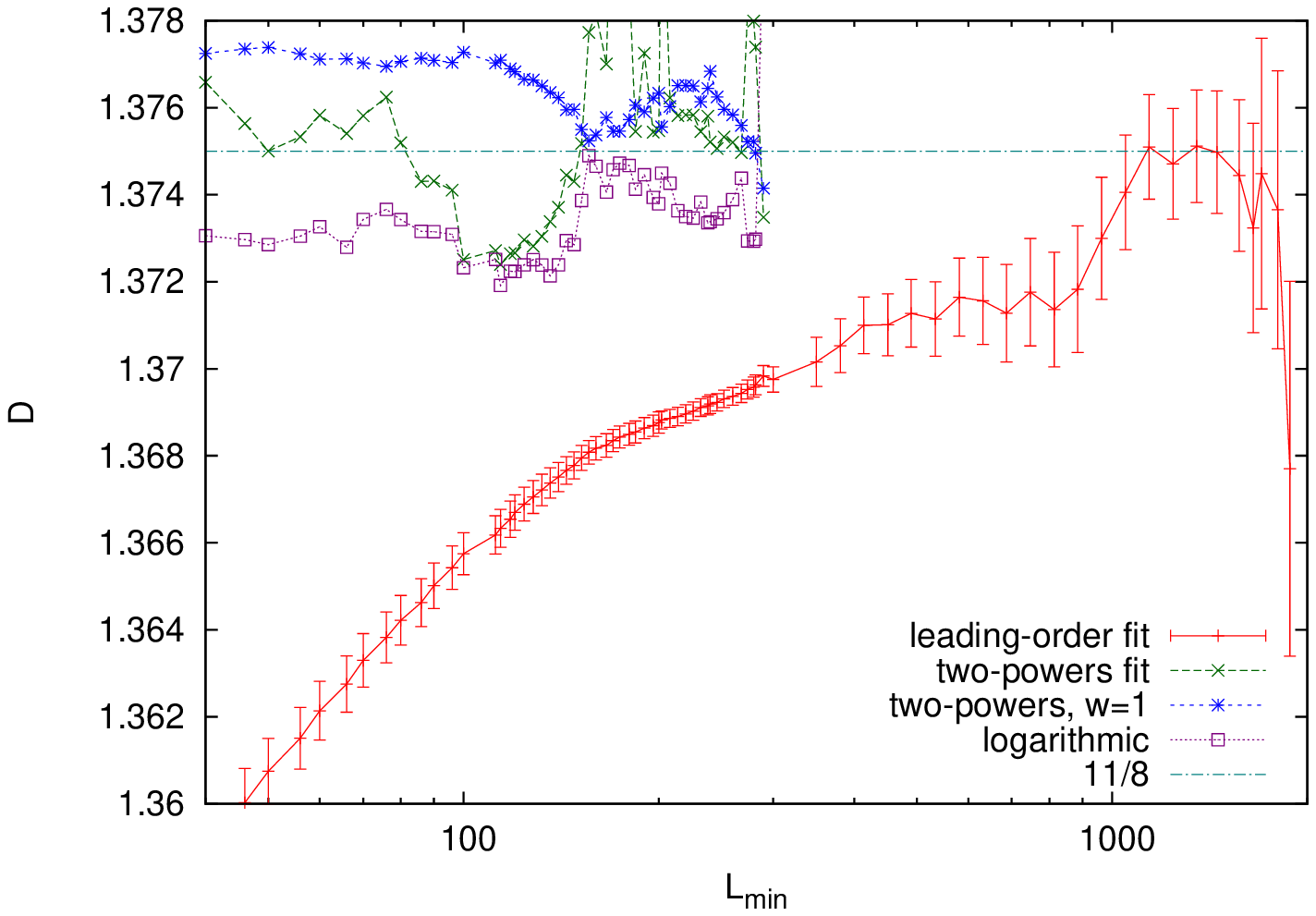}
    \caption{Values of $D$ and $D^{(1)}$ from the fits described in the text to the $S(L)$ data for the Ising point.
		Error bars for the subleading-term fits are omitted for clarity, however they never exceeded $0.008$.
		The leading-order fit gives the exact known answer only with $L_{\mathrm{min}}$ beyond 1000, while the 
		other functional forms indicate a rather mutually consistent result somewhat earlier.}
    \label{fig:ising_df_fits}
    \end{center}
\end{figure}

We noticed that the introduction of the secondary term
in the behavior of $S(L)$ makes the resulting fractal dimension generally higher than the value 
$D^{(1)}$ coming from the fit to (\ref{eq:fracdim_order1}); moreover,
the different forms for the subleading correction turn out to yield more or less compatible
values for $D$ (figure~\ref{fig:ising_df_fits}): thus, we took the spread in the pool of 
results as systematic uncertainty. In this way, we identified the fractal dimensions 
given in table \ref{table:df}.

\begin{table}
\caption{\label{table:df} Fractal dimensions obtained for each point from subleading-order and leading-order fits, as 
described in the text. Also information on the data sets is provided.}
\begin{indented}
\item[]\begin{tabular}{@{}lllll}
\br
	Point name          & $L$ values & $L$ maximum & $D$ & $D^{(1)}$ \\
\mr
	4P$_{1|234}$ 	&  25 & \phantom{0}768 & $1.4330(\phantom{0}40)$            & $1.4325(25)$ \\
\hline
	4P$_{12|34}$ 	&  36 & 1400 & $1.4805(\phantom{0}10)$            & $1.4745(15)$ \\
	D 				&  41 & 1000 & $1.4900(100)$           & $1.4720(20)$ \\
	FZ				& 115 & 2400 & $1.4320(\phantom{0}50)$ & $1.4215(15)$ \\
	B 				&  44 & 2200 & $1.3950(100)$ & $1.3915(20)$ \\
	C 				&  38 & 1600 & $1.3870(\phantom{0}90)$ & $1.3745(15)$ \\
	I 				&  79 & 2400 & $1.3751(\phantom{0}24)$ & $1.3714(\phantom{0}8)$ \\
\br
\end{tabular}
\end{indented}
\end{table}

We have observed that, within the accuracy permitted by the data (interface lengths are not self-averaging,
thus making it very difficult to get very small errors at large systems), the exponent $w$ for the power-law
secondary term is never far from one, which motivated the attemp to fit with $w$ fixed to 1.

\begin{figure}
    \begin{center}
    \includegraphics[height=2.6cm]{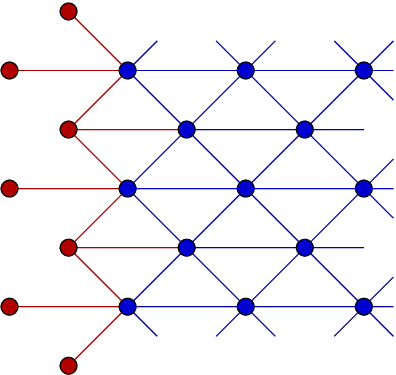}
	\hspace{0.5cm}
    \includegraphics[height=2.6cm]{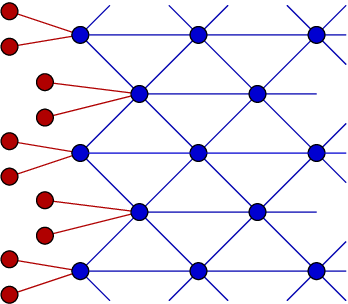}
    \caption{In the ``jagged'' way of imposing boundary conditions (left), some 
	active boundary spins (blue) connect to one fixed site 
	(red) and some other to three. In the ``non-jagged'' scheme (right), every 
	bordering active site touches exactly two fixed spins.}
    \label{fig:jaggedness}
    \end{center}
\end{figure}

A possible cause of bias in the fractal dimension could come from the influence of the fixed borders:
that is, on systems with aspect ratio of one and forced boundaries, the value of $S(L)$ is still influenced in 
a noticeable way the particular way the boundary conditions are imposed. To quantify this effect,
we have carried on two sets of simulations with same statistics and system sizes at the Ising point:
the first set (``jagged'') had the layer of fixed spins on the boundaries arranged in a way compliant
with the regular geometry of the triangular-lattice system (as all other simulations we performed),
while in the other (``non-jagged'') every
bordering site had links to exactly two fixed spins regardless of the underlying geometry (figure~\ref{fig:jaggedness}).
The resulting fractal dimensions (obtained with a leading-order fit) differ by about two standard deviations,
signalling that results from square systems are noticeably influenced by the borders.

\section{\label{sec:schramm}Crossing probability and $\SLE_\kappa$}
The analysis of the crossing probability was carried on at the same 
points where we studied $S(L)$, including the $1|234$ case for 
the Potts point.
We considered systems with aspect ratio ranging from $\ell=L_y/L_x=1$ to $5$,
and transverse side up to $L_x=300$; we collected data from about
$1$--$6\cdot 10^{5}$ configurations.

\begin{figure}
    \begin{center}
    \includegraphics[width=5.26cm,angle=270]{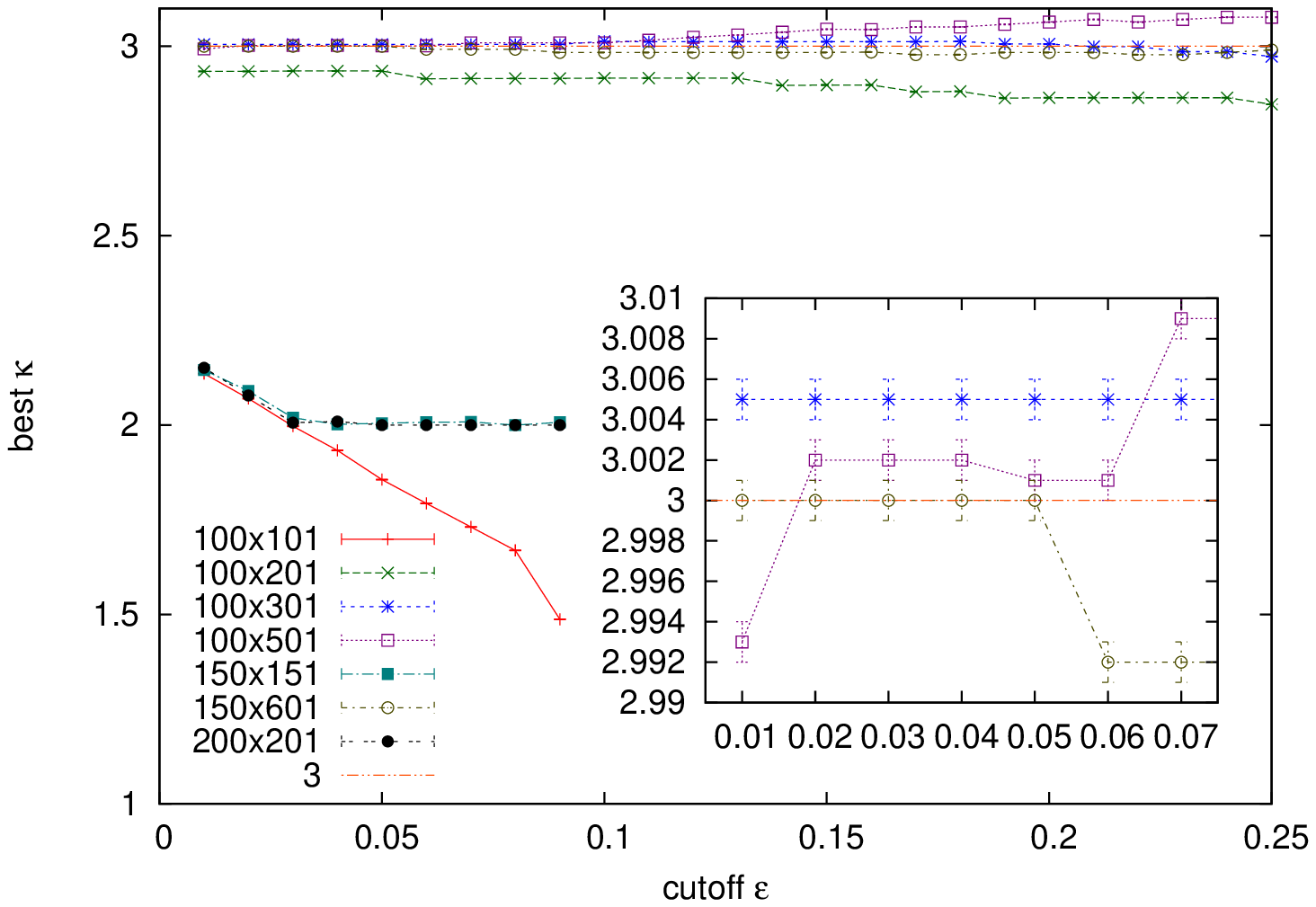}
    \caption{Best $\kappa$ as a function of the cutoff for the $12|34$ crossing probability at the Ising point
	for a variety of systems of size $L_x \times L_y$. Data start to stay constant for aspect ratios of at least three.}
    \label{fig:ising_k}
    \end{center}
\end{figure}

The basic idea, not dissimilar from the approach found, e.~g., in 
\cite{gamsacardy07, chatelain_10}, is to identify the optimal $\kappa$,
that is, the value for which (\ref{eq:starting_schramm}) best describes the
measured data, by minimising the sum of the squared deviations.
We added, however, key features to this approach, in an attempt
to keep systematic errors -- coming from the fact that the system
is discrete, finite and with limited aspect ratio -- under control:
first, we exclude from our analysis a variable halo around the endpoints,
by performing the minimisation in the domain $\epsilon<x<1-\epsilon$; this
gives the optimum as a function of the cutoff, $\kappa(\epsilon)$; 
secondly, and most important, we allow for some ``shrinking''
of the curve: this is motivated by the fact that
at the endpoints of the $x$ range the fixed boundaries squeeze, to an unknown extent,
the area where the interface can actually live undisturbed. The analysis is then carried 
on after the transformation
\eq
	x \to x' = \frac{1}{2} + r \Big( x - \frac{1}{2} \Big)\;\;,
\qe
and the shrinking factor $r$ is minimized together with the drift $\kappa$ (both found as functions of 
$\epsilon$: finally, we look for stable plateaux in $\epsilon$, at cutoffs as small as possible).
On general grounds, we expect the optimal solution $(\kappa,r)$ to have $r\lesssim 1$.
We test this analysis at the Ising point, where it is known that the $12|34$ interface
is described by a $\SLE_\kappa$ with $\kappa=3$, in accordance with our $S(L)$ investigation:
the result, for elongated enough systems, is $\kappa=3.002(3)$ and $r=0.98$ (figure~\ref{fig:ising_k}), to be compared
with the outcome of the basic analysis on the same data, $\kappa=3.08$ (the same value found in \cite{gamsacardy07}).

\begin{figure}
    \begin{center}
    \includegraphics[width=5.26cm,angle=270]{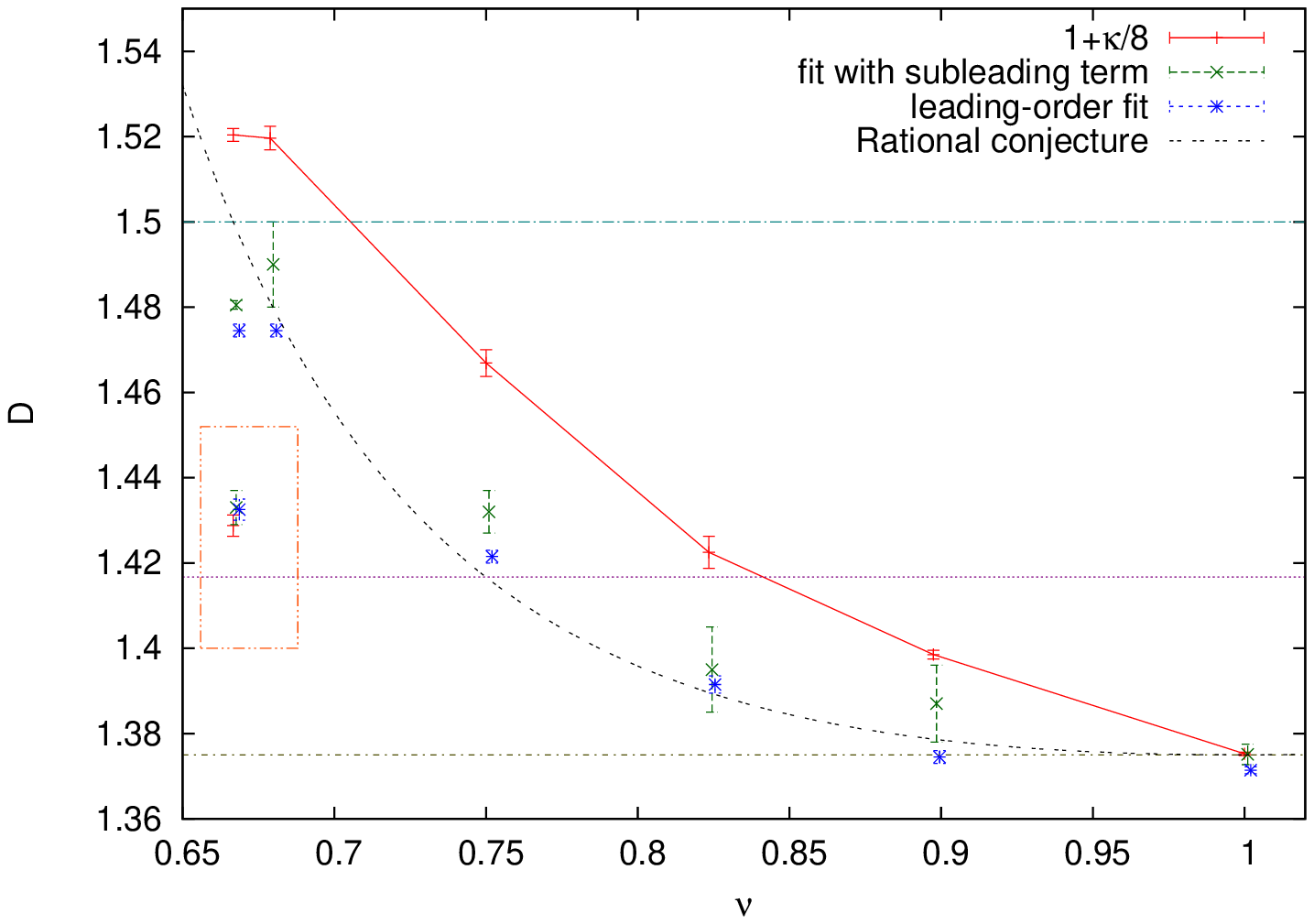}
    \caption{Comparison between fractal dimensions coming from various fits to $S(L)$ (datapoints),
    from the $\SLE$ relation $D=1+\kappa/8$ (red line with error bars), and the polynomial Ansatz 
    (\ref{eq:polynomial_ansatz_d}) (smooth line).
    The rectangle on the left encloses results for the $1|234$ interface, all other points refer to the $12|34$.}
    \label{fig:total_df_plot}
    \end{center}
\end{figure}

We then proceeded to apply this technique to the other points: the shrinking factor 
was around $0.96$--$0.98$ for all $12|34$ interfaces, while the $1|234$ at the Potts
point yielded the puzzling result of $r\sim 1.025$; anyway, we identified a stable
value of $\kappa$ for all points considered. In all cases, we start finding plateaux
and mutually consistent values only at aspect ratios of $\ell=3$ or more.
The values of $\kappa$ can be compared with the relation between drift and interface 
fractal dimension from $\SLE$, $D=1+\kappa/8$, and with the theoretical prediction 
at the exactly-known points, see table \ref{table:kappa} and figure~\ref{fig:total_df_plot}.

\begin{table}
\caption{\label{table:kappa} Comparison, for all points studied, between the numerical drift $\kappa$
obtained from the Schramm formula and the corresponding value as extracted from the interface fractal
dimension. Also, when available, the theoretical prediction is provided.}
\begin{indented}
\item[]\begin{tabular}{@{}llll}
\br
	Point name & numerical $\kappa$ & $8(D-1)$ & theoretical $\kappa$ \\
\mr
	4P$_{1|234}$ 	& $3.430(20)$ & $3.464(32)$ & $4$ \\
\hline
	4P$_{12|34}$ 	& $4.163(12)$ & $3.844(\phantom{0}8)$ & $4$ \\
	D				& $4.157(22)$ & $3.920(80)$ & -- \\
	FZ				& $3.735(25)$ & $3.456(40)$ & $10/3$ \\
	B 				& $3.380(30)$ & $3.160(80)$ & -- \\
	C 				& $3.188(\phantom{0}8)$ & $3.096(72)$ & -- \\
	I 				& $3.002(\phantom{0}3)$ & $3.001(19)$ & $3$ \\
\br
\end{tabular}
\end{indented}
\end{table}

The $12|34$ interface gives a drift compatible with the fractal
dimension only at the Ising point, suggesting that as one moves away from
Ising the interface is no more described by a simple $\SLE_\kappa$; moreover,
at the Potts point we observe compatibility for the $1|234$ interface, but not
in agreement with the theoretical expectation $\kappa=4$.

\section{\label{sec:conclusions} Discussion and conclusions}
We studied the particular interfaces $12|34$ in the Ashkin-Teller model. The fractal dimensions were 
calculated all along the critical line in the interval between the Ising model and the four-states 
Potts model. 
We also checked Schramm's formula to verify whether our interfaces are $\SLE_{\kappa}$ or not. Our 
numerical calculation shows that the interface $12|34$ is not $\SLE_{\kappa}$ except at the Ising 
point. This may be expected from lattice arguments: our interfaces do not exhibit Domain 
Markov Property, at the lattice level, except at the Ising point; consequently, the property does not 
hold at the continuum level as well. However, this argument is not always true: it is possible to 
have no domain Markov property at the lattice level -- as it seems the case for the disorderd systems -- 
but to recover it in the continuum limit \cite{BDM}. Our numerical calculation shows that most probabily this 
phenomenon does not occur in our system. That means, even if the interfaces were 
conformally invariant we can not expect a match between the results coming from the fractal 
dimension calculations and Schramm's formula. Although our numerical calculation shows that the 
interfaces are not $\SLE_{\kappa}$ they can still be conformally invariant and related to some 
generalizations of $\SLE_{\kappa}$, such as $\SLE_{\kappa,\rho,\rho}$. 
One can use the above argument also for the $1|234$ interfaces with the difference that, in this 
case, we have Domain Markov Property just at the four-states Potts point. Our numerical calculation 
shows that the results coming from calculating the fractal dimension and Schramm's formula 
agree; however, the result is not compatible with $\SLE_{4}$.
The mismatch can be related to the expected very slow convergence to the infinite-volume
regime, with the appearance of logarithmic corrections (see for example \cite{asikainen_et_al_potts_fss,
adams_et_al_potts_fss,zatelepin_et_al_potts_fss}),
however our results do not show a significant trend towards $\kappa=4$ as the system is enlarged.
At the FZ point, our calculation shows that, although the numerical outcome for the fractal dimension 
is almost compatible with the prediction in \cite{Santachiara}, there is still a discrepancy 
that did not seem to vanish even when simulating to the very huge system sizes.

Accepting the conjectures at the Potts point $D=\frac{3}{2}$ and the FZ point $D=\frac{17}{12}$, and 
considering the exact result at the Ising point $D=\frac{11}{8}$, one can put forward an Ansatz about the fractal 
dimension of the $12|34$ interfaces as follows: since the critical exponents in the Ashkin-Teller 
model are related to the compactification radius of the orbifold with a rational function, it is plausible 
to argue that the relation between $D$ and the compactification radius is rational as well. Using the three 
known points one can derive the following equation:
\eq
	D = \frac{7}{8} + \frac{r^2}{8} + \frac{1}{2r^2}\;\;,
	\label{eq:polynomial_ansatz_d}
\qe
where $r$ is the compactification radius of the associated bosonic theory. The number 
$\frac{7}{8}$ appearing there could be a signal of the important r\^ole of the operator 
$\tau$ in calculating the fractal dimension of our interfaces.
By combining (\ref{AT_thermal_exponent}) with the relations, valid for the 
Ashkin-Teller model \cite{mohammad_steo_isoradial},
\eq
	\nu(\lambda)=\frac{2\pi-2\lambda}{3\pi-4\lambda}\;\;\;;
	\;\;\;\lambda(\beta)=3\arcsin\frac{\sqrt{3-\frac{1}{\tanh 2\beta}}}{2}\;\;
\qe 
(the second one is specialized to the triangular lattice), we find the following expression, 
useful for checking the above Ansatz:
\eq
	r^2(\beta) = 4 - \frac{12}{\pi}\arcsin\Bigg( \frac{\sqrt{3-\frac{1}{\tanh(2\beta)}}}{2} \Bigg)\;\;.
\qe
In figure~\ref{fig:total_df_plot} we have compared this prediction with the results coming from the 
numerical calculation. The formula roughly describes the numerical data, but with no perfect compatibility.
An important feature of this formula is the prediction of a minimum for $D$ at the Ising point,
that was recently confirmed numerically in \cite{ps3}. In addition, the above Ansatz predicts the correct
value $D=\frac{3}{2}$ for the $XY$ point. Interestingly, 
one can map the model at the $XY$ point
into the $O(n=2)$ model on the honeycomb lattice.
The argument goes as follows: at the $XY$ point of the triangular lattice, the weight associated to neighbouring 
sites with different 
spins, i.~e.~$W(\tau_{i}\neq\tau_{j}, \sigma_{i}\neq \sigma_{j} )$, is zero. This can happen just 
at the $XY$ point of the triangular Ashkin-Teller model because there we have  
$1-2x_{1}+x_{2}=0$.
Renormalizing the weight of the neighbouring interactions with like spins to one, 
i.~e.~$W(\tau_{i}=\tau_{j}, \sigma_{i}= \sigma_{j} )=1$, gives 
$W(\tau_{i}\neq \tau_{j}, \sigma_{i}= \sigma_{j} )= W(\tau_{i}=\tau_{j}, \sigma_{i}\neq \sigma_{j} )=\frac{1}{\sqrt{2}}$.
If we draw a line on the honeycomb-lattice link dual to the $(i,j)$ link of the original lattice for all
pairs of unlike spins $\sigma_{i}\neq \sigma_{j}$, we will get a loop model with the 
following partition function:
\eq
         Z = \sum_{\cal C}\Big(\frac{1}{\sqrt{2}}\Big)^{b}2^{d}\;\;,
\qe
where $b$ is the number of bonds, $d$ is the number of loops and the sum is over all loop configurations ${\cal C}$. 
The above formula is the partition function of the $O(n=2)$ model at the critical point \cite{nienhuis2}. 
It was conjectured by many authors, see for example \cite{Kager}, that these interfaces are described by $\SLE_{4}$. 
This result is even stronger than the prediction of our formula because it also claims that the interfaces 
are conformal and related to the simple $\SLE_{4}$. The very interesting property of the $XY$ point is that 
most of the interfaces that can be defined \cite{ps3}, including $1|234$, exhibit Domain Markov 
Property: thus we conjecture that they are all $\SLE_{4}$. 

Another very interesting phenomena is that for the Ashkin-Teller model we
expect different continuum limits for the same boundary conditions on different underlying lattices. 
The simple way to see this is to consider the end point of the physical portion of the critical line on the square 
lattice, which is $(x_{1},x_{2})=(\frac{1}{2},0)$ with $\nu=2$. Here we expect the same behaviour 
as the $XY$ point of the triangular lattice, i.~e.~the weight associated to neighbouring
sites with different spins is zero and so we expect to have Domain Markov Property. It means that if the 
interfaces were conformal we would expect simple $\SLE_{\kappa}$. This is not necessarily true on the triangular 
lattice with $\nu=2$ because at the lattice level we do not have Domain Markov Property. The conclusion is 
the same boundary conditions on the different lattices for the statistical models with some internal symmetries 
could have different continuum limits.

Finally, we would like to make some observations on the numerical pitfalls in identifying fractal dimensions
correctly. The triangular lattice has the advantage of allowing for a unambiguous interface definition,
i.~e.~with no need to define a ``tie-break'' rule to deal with four-lines junctions as is the case for the
square lattice; on the other hand, it seems that this geometry somewhat enhances the finite-size effects,
requiring simulating very large systems to achieve a stable value for $D$ (as opposed, for instance, 
to the rather stable results from moderate sizes for the square lattice \cite{ps3}). 
Moreover, even though we could not apply the predictions in \cite{Aharony_Asikainen_analytical_corrections}
for the subleading terms in $S(L)$ (because they are formulated in a different setting than ours), there are some indications
that fitting numerical data simply to the leading-order behaviour does not give reliable fractal dimensions.
Another potential concern is our choice -- widely common in literature -- of working with 
square aspect ratios $\ell=1$: as shown by the test of jagged vs.~non-jagged boundary 
conditions, a proper determination of $D$ should reasonably be conducted on systems fulfilling 
to some extent the requirement $L_x \gg L_y$.

\ack
We are indebted to F.~Gliozzi for fruitful discussions and to M.~Picco 
and R.~Santachiara for sharing various numerical results.
M.~A.~R.~thanks M.~Ghaseminezhadhaghighi for early discussion of this problem.
S.~L.~wishes to thankfully acknowledge the Center for Scientific Computing 
of Frankfurt and the Universit\`a di Torino for providing the 
computational resources. Most of the work of M. A. Rajabpour was done when he 
was at the Universit\`a di Torino.

\end{document}